\documentclass{article}
\usepackage[active]{srcltx}
\usepackage{ amsthm, amssymb,amsfonts}
\usepackage{graphicx}
\usepackage{epsfig}
\usepackage{latexsym}
\usepackage{colortbl}
\usepackage{times}
\usepackage{helvet}
\usepackage{courier}
\usepackage{caption}
\usepackage{subfig}
\usepackage{float}
\usepackage{xcolor}
\usepackage{pslatex}
\usepackage{eso-pic}
\usepackage[bottom]{footmisc}
\usepackage{url}
\usepackage{etex}
\usepackage{authblk}
\begin{document}

\title{Accountability in AI: From Principles to Industry-specific Accreditation}

\author[1]{Chris Percy}
\author[1]{Simo Dragicevic}

\author[1]{\\Sanjoy Sarkar}
\author[2]{Artur S. d'Avila Garcez}
\affil[1]{Playtech plc, London, UK, e-mails: \{Christian.Percy, Simo.Dragicevic, Sanjoy.Sarkar\}@playtech.com}
\affil[2]{City, University of London, UK, e-mail: a.garcez@city.ac.uk}
\maketitle
\date{}                 
\setcounter{Maxaffil}{0}
\renewcommand\Affilfont{\itshape\small}

\begin{abstract}
Recent AI-related scandals have shed a spotlight on accountability in AI, with increasing public interest and concern. This paper draws on literature from public policy and governance to make two contributions. First, we propose an AI accountability ecosystem as a useful lens on the system, with different stakeholders requiring and contributing to specific accountability mechanisms. We argue that the present ecosystem is unbalanced, with a need for improved transparency via AI explainability and adequate documentation and process formalisation to support internal audit, leading up eventually to external accreditation processes. Second, we use a case study in the gambling sector to illustrate in a subset of the overall ecosystem the need for industry-specific accountability principles and processes. We define and evaluate critically the implementation of key accountability principles in the gambling industry, namely addressing algorithmic bias and model explainability, before concluding and discussing directions for future work based on our findings. \textbf{Keywords:} Accountability, Explainable AI, Algorithmic Bias, Regulation. 
\end{abstract}

\section{Introduction}

Since Isaac Azimov’s “Three Laws of Robotics” in the 1940s [1], discussions of advanced and increasingly autonomous technology have gone hand in hand with concerns over control and accountability. More recently, the concerns around AI have turned away from the prospect of a killer robot and towards AI systems already in deployment today which can automate inequality [2], with potential for inflicting a life-or-death impact through an automated decision-making. Recent AI-related scandals involving leading companies have shed a spotlight on accountability in AI [3]. The arch futurism that characterised the early philosophy and ethics of AI accountability has given ground to the prosaic pragmatism of EU Expert Groups, White Papers, and regulatory debate [4]. 

As of late 2020, some 100 sets of published ethical principles to govern AI had been identified by researchers, with the majority published since 2016 [5]. However, publishing documented principles alone has proved insufficient to convince external stakeholders in civil society or the wider public that a company or research team will uphold those principles. Given imperfect alignment of interests among market economy players and the imbalances in power, access to inside information and technical understanding, such limited trust is hardly surprising. As such, there is a growing demand for high-level ethical principles to be translated into specific practices and accountability mechanisms. For example, Stanford University’s 2021 AI Index Report identifies a shortage of benchmarks in AI ethics as one of its top nine takeaway industry trends [6]. In 2020, a need for better regulatory standards and quality assurance was identified by a UK Government commission as one of the five key trust-related barriers holding back AI [7]. AI Communications has recently published a game theoretic approach to the value alignment problem applied to adopters and non-adopters of AI technology, concluding that external regulation is needed if the gains for the adopters are not to be achieved at a cost for non-adopters [8].

Governments have been responding to a high tempo of public concern around AI by exploring regulation. The European Union published draft regulatory proposals in April 2021 [9], with the debate progressing twice as fast towards regulation as the earlier debate on data protection, and with the risk that alternatives to regulation may not be given full consideration [10]. At the same time, a research literature has been emerging on non-regulatory mechanisms by which AI teams might be held accountable to ethical principles, whether set by the team itself or by an external body. Most of these are focused on the need for accountability in general, or identify a specific mechanism such as tools for developers [11], risk assessment [12], translation into design principles [13], a focus on the tensions that may arise when trying to implement principles [14], or algorithm audits [15]. Such mechanisms and tools are promising but remain immature, with significant extra development required before they might mitigate meaningfully the concerns of civil society.

In this paper, AI is defined broadly as spanning the technologies and functions addressed in the Stanford University's AI Index [6]. Our primary interest is in AI as decision-maker or decision-informer (e.g. Machine Learning (ML) models, autonomous systems). The misuse of personal data is kept as a separate ethical issue, albeit one that seems to be contravened by the same actors as those working on ML for decision-making. We agree with researchers who note that precise definitions become important for the design of specific accountability mechanisms [16]. Accountability is also defined broadly in this paper, including both narrow mechanisms defined in terms of Bovens’ framework [17] and softer forms of leverage or influence.

We argue that accountability for AI as a set of emerging technologies should be thought of in terms of an accountability ecosystem in which specific mechanisms (such as those mentioned in [11]-[15]) are individual contributory elements. At present the ecosystem is unbalanced, which can be seen in the failures of certain mechanisms that have been attempted by leading technology companies. By taking an ecosystem perspective, we can identify certain elements that need developing and bolstering in order for the system as a whole to function effectively. Corporate governance mechanisms such as standardised processes and internal audit frameworks, leading up to potential external accreditation, need to be made to work together in ways that go beyond regulatory requirements, especially in technologies' early period of evolution and deployment when regulation lags practice. 

Specifically, this paper brings to life the notion of accountability ecosystem through an industry-led programme of work to implement industry-specific accountability in AI, as called for in [12]. As part of the Research and Development (R\&D) programme at Playtech plc, a global software provider and operator, operating primarily in the gambling sector with a focus on online gambling, the paper reports the use of AI systems to help reduce harm from gambling. We identify the use of AI for responsible gambling as a relevant test case for industry-specific accountability operationalisation, due to a high regulatory focus, divergent regulatory perspectives worldwide, and longstanding debate over other ethical dilemmas including addiction. We relate results obtained from the risk profiling of gambling behaviour using ML and explainability [18] and lessons learned from problem gambling classification, indirect bias and the need for algorithmic fairness [19] to an accountability ecosystem and its operationalisation. Two key elements of the accountability ecosystem are discussed in detail: interventions to reduce bias and increased transparency via model explainability. The benefits of an industry-specific accountability process are illustrated in that it can be documented, reviewed, benchmarked, challenged and improved upon, both to build trust that the underlying ethical principle is being taken seriously and to identify specific areas to do more. 

In a nutshell, on ML algorithmic bias, we study gender bias when applied to model accuracy, identifying pro-female bias in one gambling operator model. A gender-blind intervention technique is developed that reduces gender difference in True Positive Rate by 44\%, albeit with some cost to overall accuracy. On ML transparency and black-box system explainability, we study feature-specific global model explanations which produce so-called feature risk curves. These have supported the identification of blind spots in the model and a redirection in the programme of R\&D. We explain the role of the above process within a broad accountability ecosystem and reflect on the work required in future to address the ethics of AI as a whole. Our conclusions support the importance of industry-specific approaches to the operationalisation of accountability principles, noting how different metrics, priorities and accountability processes arise in online gambling compared to what might arise in other industries.

The remainder of the paper is organised as follows. In the next section, we place the contribution of the paper in the context of the demands for accountability, the current discussion around AI regulation, and the prevailing approaches to accountability today. We then present the accountability ecosystem perspective and suggested areas of development for the sector as a whole. We then present and discuss the Playtech experience and results on bias and transparency before listing the proposed industry-specific principles and processes. We conclude with a critical evaluation leading to a brief discussion on directions for future work. 

\section{Background and Related Work}

\subsection{Demands for AI accountability}

Outrage on AI accountability has been sparked by diverse issues, including racial bias in US healthcare algorithms [20], college admissions [21], image classification accuracy [22], and the justice system [23], as well as public protests around algorithms more generally, such as lack of algorithmic transparency in welfare fraud detection in the Netherlands in 2018 [24] or concern over skew in exam grading in the UK in 2020 [25]. The recent documentary, The Social Dilemma, and a U.S. Senate hearing of the Commerce, Science and Transportation Committee, \emph{Testimony from a Facebook whistleblower},\footnote{Protecting Kids Online: Testimony from a Facebook Whistleblower. U.S. Senate Subcommittee on Consumer Protection, Product Safety, and Data Security Hearing. https://www.commerce.senate.gov/2021/10/protecting\%20kids\%20 online:\%20testimony\%20from\%20a\%20facebook\%20whistleblower, 5 Oct 2021.} also highlight serious concerns of AI’s influence on topics such as the mental health of youth exposed to social media platforms, and the use of private data to manipulate information leading up to addiction to the social media platforms. In the gambling sector, the Finnish monopoly operator Veikkaus set out five guidelines for ethical AI in its gambling operations in January 2021 [26]; concerns have been raised by sector leaders regarding model opaqueness and quality, systemic bias, and its elevation of rules and quantitative metrics ahead of human interactions [27].

The annual AI Now reports from New York University enumerate various scandals, such as in 2019 the high carbon cost of training large-scale models, the biases encoded in inferring emotions from micro-expressions, and worker exhaustion from gig economy productivity algorithms [28]; and in 2018 the outrage around algorithmic advertisement targeting for US and UK elections with Facebook and Cambridge Analytica, unsafe and incorrect cancer treatment recommendations from IBM Watson, and deaths caused by autonomous cars [29]. These recent scandals take place alongside ongoing, long-standing concerns about long-term existential risk from AI and the need for careful governance, e.g. the 2015 AI Open Letter from the Future of Life Institute [32], Elon Musk’s call for more regulation in 2017 [30], and Nick Bostrom’s framing of the AI control problem [31]. 

Public opinion surveys in the US [33] show broad agreement that AI needs to be carefully managed (84\% in 2018), with increased regulation for tech companies (51\% in 2018 versus 9\% wanting less regulation), although with disagreement on what this means in practice and with low levels of trust in government. Desire for governance is greatest around data privacy, AI-enhanced cyber attacks, surveillance and digital manipulation, as well as autonomous weapons, vehicles and value alignment [34]. A 2019 poll for the UK Government’s Committee on Standards in Public Life identified that a net 22\% of respondents were not confident that government would use AI ethically, with confidence most likely to be increased by a human operator having the final say, a right of appeal to a human operator, and confirmed levels of accuracy with easy-to-understand explanations for AI decisions [35]. A comparable level of concern emerges globally, with a 2019 Ipsos poll of 20,107 adults from 27 countries revealing 48\% wanting greater regulation of corporate AI tech (vs 20\% disagreeing) [36]. 

Major global civil rights groups have set up monitoring units for technology and AI-related topics, such as Human Rights Watch [37] and Amnesty [38], both reflecting concerns about the ethical implications of AI in the present, and providing further investigative resource and dissemination effort to increase awareness of future scandals.

\subsection{Rapid progression towards regulatory mechanisms}

Spurred in part by the growing scandals and public concern, regulatory activities have accelerated in recent years. In particular, the journey from general ethical principles to regulation appears to be happening much faster for AI than for data protection, suggesting heightened risks of ineffective regulation. 

Concerning the data protection debate in the late 20th century, the OECD first issued principles for data protection as recommendations to the European Council in 1980 [39], being endorsed (but not legislated for) by the US government. It would not be until fifteen years later in 1995 that the European Union introduced the Data Protection Directive [40], implemented from 1998. This Directive has now been superseded by the much stronger General Data Protection Regulation (GDPR), the development process for which began in 2012, with the new regulations enforceable from 2018 [41]. GDPR is proving an influential model for regulations in other countries such as Brazil, China, Japan and some USA states (e.g. the California Consumer Protection Act) [42]. 

Concerning AI, by contrast, the EU published its proposals for regulating AI in April 2021, with the OECD’s non-legally binding principles on Artificial Intelligence published and widely adopted in 2019 [43]. A literature review in 2019 identified 84 published sets of ethical principles or guidelines, with the first focused on robotics in 2011 (from the UK’s EPSRC), but with 88\% released after 2016 [44]. Even if the OECD were a late mover on data protection but an early mover on AI regulation, the rate of progression from widespread discussion of principles to EU regulatory proposals appears to be twice as fast for AI than data protection. With government-level discussions related to AI regulation, standardisation or governance launched at the G20 [45], the United Nations [46], China [47] and Russia [48], it is possible that regulation may materially and forcefully direct AI activity in the near future.

Early regulation may be an asset or a liability for AI as a whole, and we may never be truly sure which. There are often-cited (albeit always contested) examples on all sides of the debate. Early regulation is argued to have quashed innovation to the likely detriment of public well-being (e.g. the delays to the early automotive industry [49]) or entrenched large corporate incumbents who are able to service the cost of regulatory compliance (e.g. pollution regulation [50]). In other circumstances, under-powered early regulation may have provided a false sense of security before a new wave of scandals turns the public against the technology even once better safeguards are in place (e.g. the accidents that derailed nuclear industry development [51]). Regulation can also be too late, with proliferating scandals leading to a likely delay in beneficial progress or direct harm (e.g. lack of regulation of radioactive drinks as popular consumables in the 19th and 20th century [52]). A patchwork approach can also lead to global disparities in regulation, producing hot-spots of ethical failings, risky behaviours and perhaps game-changing innovation; short-term regulatory divergence might also permit a productive experimentation with different techniques, as discussed by [16] as a ‘race to AI regulation’.  

Governments are well aware of the balancing act between free innovation and ethical caution, between fostering funding environments alongside public trust, and between local leadership vs collective action. But neither Governments nor any other participant is truly confident – at least not convincingly so - in how society and technology might evolve. Even if outcomes were well-known, there are differing value palettes against which those outcomes would be assessed. However, there is a tendency, noted by some government bodies and researchers, for attention to be focused on formal regulatory mechanisms such that other options are not properly considered. Better cost-benefit outcomes are sometimes missed, giving rise to a literature aimed at government decision-makers called ‘alternatives to regulation’ [53]. This motivates adopting a broad perspective in understanding the different mechanisms that can enforce accountability and how they relate.

\subsection{Accountability in AI}

We now describe three key aspects of the prevailing approach to corporate accountability for AI in recent years, considering both major technology companies and major research institutes focusing on AI. First, many actors have focused their public-facing work on developing sets of ethical principles. Second, leading tech companies, such as Google and Facebook, have also begun exploring efforts to engage external experts, whether to lead internal initiatives, sit on boards or publish audits, but faced significant difficulties in making these initiatives gain positive traction. Third, there is an emerging discussion of additional mechanisms to support accountability, unsurprising given the scale and diversity of the field, but these have not yet become dominant in the debate. Collectively, these efforts appear to have been insufficient to date to mollify the concerns of civil society or inform the rapid progression towards regulation mentioned above.

A review in 2019 identified 84 published sets of ethical principles or guidelines [44], with a short ACM Viewpoint paper in December 2020 noting that there were now over 100 such sets of principles (with more being added regularly) [5]. These principles are produced by diverse AI sector actors, with the 2019 review identifying 23\% as private companies, 21\% as government agencies, 11\% as research institutions, 10\% as international bodies, 8\% as scientific societies, 8\% as non-governmental organization (NGOs), with the remainder made up of private sector, research alliances, unions and political parties. Several examples were sourced from each of Japan, Germany, France and Finland, with several other countries having one example, but most documents were drawn from the USA (24\%) and the UK (17\%). 

Researchers have pointed to substantial overlapping [14] and an emerging, if imperfect consensus [44] between the published principles, particularly with regard to transparency, justice and fairness, non-maleficence, responsibility and privacy. Nonetheless, researchers have also called for a greater level of detail regarding the need to operationalise the principles [5, 12], noting that different groups may interpret the phrasing of principles differently and that high-level principles tend to elide rather than help resolve tensions.\footnote{E.g. between meaningful consent over personal data vs rapidly building more accurate models; algorithmic decision-making (and sometimes accuracy) vs fair treatment; personalisation vs citizenship and solidarity; automation to support convenience vs self-actualisation and dignity [14].} It is possible that ethical principles may yet drive a sufficient change in behaviour to satisfy ethical concerns, but with a proliferation of principles since 2016, it is unlikely that volume alone will be fast enough, at least not without other supporting measures. 

Leading AI companies are additionally investing in internal groups, processes and research in an effort to demonstrate a commitment to ethical behaviour. Success here too appears limited, at least to date. For instance, Google cancelled its advisory ethics board in 2019 (Advanced Technology External Advisory Council) following concerns over certain board members’ positions on civil rights issues and their military connections [54]. In December 2020, Timnit Gebru ceased to work at Google, having been the technical co-lead of their Ethical Artificial Intelligence Team, in a series of high-profile disagreements which saw Gebru describe the company as institutionally racist [55]. A less publicly controversial angle for the technology firms has been to provide tools to help the development community analyse and understand their algorithms, e.g. Microsoft Interpret, Google’s open source What If tool (launched in TensorBoard in 2018) and Facebook’s internal Fairness Flow tool, tested in 2018. However, such services have done little to abate criticism of the companies’ own applications of AI.

Facebook cooperated with a civil rights audit from 2018 to 2020, led by Laura Murphy and covering key issues in the core Facebook app such as election influence, content moderation, diversity and inclusion, advertising, algorithmic bias and privacy [56]. 
%This external review noted a number of positives in the access to the company but equally drew attention in its summary to recent “vexing and heart-breaking decisions” (p8) by Facebook that had “real world consequences that are serious setbacks for civil rights” (p6) and described Facebook’s overall approach as “too reactive and piecemeal” (p8). 
The audit was the first of its kind and faced process challenges in terms of scope and benchmarking (the audit explicitly does not compare Facebook against other firms). Despite some positive messaging and Facebook’s cooperation, the civil society response at the time was mostly negative [57, 58]. 
%Since closing in July 2020, there has been no publicised discussion of further audits or systematising into structured, repeatable processes that might track progress or generate benchmarking insights. A new position at Facebook, VP of Civil Rights, was hired in January 2021 and marks a repeated effort at tackling the topic, launching a new internal civil rights organisation, but with no discussion of audit-type approaches or external validation [58].

There is emerging discussion of other accountability mechanisms short of direct regulation, e.g. risk management and audit processes [12], design principles [13], the introduction of reviewability and traceability requirements into model development and implementation [78, 79], and operational checklists prioritised according to value judgements on which ethical principles are most important in a particular context [5], potentially supported by internal ethics committees borrowing on approaches from biomedical research [59]. However, these have not yet become driving features of the public-facing debate. A large number of tools is potentially capable to support the operationalisation of principles, with a 2020 paper from the Oxford Internet Institute [11] identifying a list of over 100 sources with tools and techniques, ranging from checklists and ethical discussion prompts to visualisation software and peer-to-peer networks, aiming to aid Machine Learning developers to implement particular ethical principles at particular stages of software development. Tools have also been developed to help users control the results of recommendation algorithms, to surface and adjust for bias that might emerge. One such tool was presented at the IJCAI-ECAI-18 Demonstration Track for controlling polarisation in news feeds [60]. 

A quote from the AI Now 2019 report summarises the prevailing critical position: \begin{quote} \emph{Companies, governments, NGOs and academic institutions continued to dedicate enormous efforts to generating AI ethics principles and statements this year. However, the vast majority of these say very little about implementation, accountability, or how such ethics would be measured and enforced in practice} [28, p11].\end{quote}

\section{The Accountability Ecosystem perspective}

In this section, we propose an accountability ecosystem approach for AI as a useful lens for understanding how current practice might be improved. An ecosystem aims to provide a way of thinking about the diverse stakeholder roles and corresponding forms of accountability mechanisms that might collectively enable the AI industry to move past its current situation and live up to its stated principles. From this perspective, we consider that various stakeholders may wish to hold a particular actor, such as a product team drawing on AI technologies or a research unit developing new technologies, to meet certain ethical guidelines or rules. The key insight is that accountability tools work better as part of an ecosystem: using some mechanisms prematurely without the supporting infrastructure can backfire, as illustrated by Facebook's audit as discussed earlier.

The notion of an accountability ecosystem has been proposed in other diverse fields such as sexual and reproductive health and rights [61], information and communication technology [62] and government accountability [63]. The concept seeks to move beyond the common, simple formulation of "transparency + participation = accountability" into a complex formulation in which accountability operates through the "relationship between multiple levels of government, citizen collective action, civil society advocacy, and institutions, wrapped together by a web of social, political, and cultural factors in a given country context" [64]. A recent literature review on algorithmic accountability similarly emphasises challenges around networked accountability dispersed across many roles and fora, and the need to consider the entire socio-technological process within which algorithms are developed and implemented [65].  

As [64] explains, accountability systems and practices are also shaped by the same power relationships that give rise to the need for them, being the imbalances in information access, operational control, and technical understanding between the team implementing the technology and the technology's users, targets, or audiences, with trust further limited by the awareness of different incentives between the implementing team and other stakeholder layers (see Figure 1). Any single accountability mechanism is typically insufficient to overcome such distrust, especially where mechanisms with close visibility of the implementing team have likely been shaped by the team itself or those sympathetic to it - as is particularly necessary in areas of innovation where technical understanding is limited outside of the R\&D teams. An ecosystem approach of multiple, overlapping accountability systems, bridging out from the implementing team towards external audiences, can mitigate these concerns, while noting that uncertainty, risk, and distrust will continue to wax and wane in response to events, along with corresponding levels of scrutiny and restrictions.

In the gambling sector, a closely related concept was introduced by financial sector experts invited to a responsible gambling roundtable in 2016 [66]. Following the financial sector’s reckoning with its failed credit risk algorithms in the lead up to the 2008 financial crisis, there was a need to add stronger layers of accountability. As one industry leader explained [74]: “In financial services, it is typical to implement a ‘three-line of defence’ development and governance process to enable effective model oversight to safeguard against such risks. Oversight of model development and implementation is effectively triangulated by 1) internal compliance or finance teams, 2) internal audit teams, and 3) by external experts.”  

Figure 1 provides a simplified schematic for an accountability ecosystem focused on company-based AI teams in a mixed market democracy, with four illustrative \emph{stakeholder layers} in decreasing proximity from the AI technology actor being held to account. The corporate layer, corresponding to the board and senior management, seek to become confident that the actor is acting in accordance with their wishes and interests to create a profitable, secure and sustainable business. The actor’s market counterparts, such as various current or prospective investors, creditors, customers, employees and partners, wish to be confident that their trades will be honoured in good faith and not bring reputational damage. Civil society, defined broadly to include activists, NGOs, media actors, lobbyists and research groups, reflects individuals and subgroups with a set of values and preferences about how they wish society to evolve, including its underpinning technologies and AI. Finally, government and its own accountability system of parliament and electorate have a stake in AI behaviour through their desired stewardship of a thriving economy and a stable, secure society.

\begin{figure*}
  \centering \includegraphics[width=1.0\textwidth]{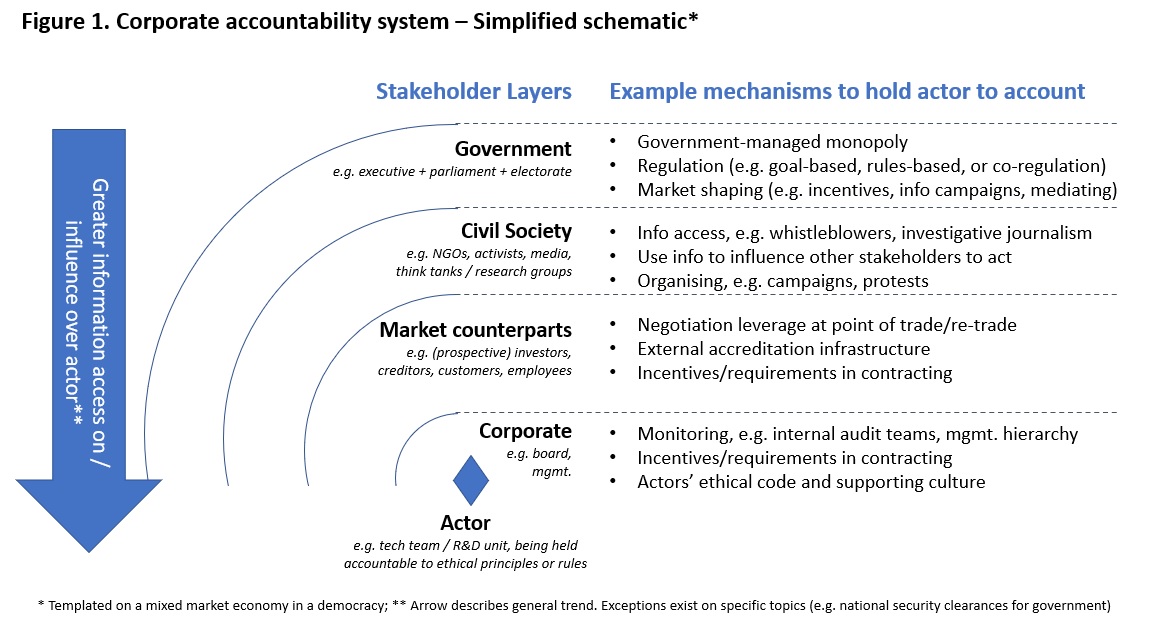}
  \caption{The Accountability Ecosystem.}
\end{figure*}

Each stakeholder layer has a different set of primary accountability mechanisms. The strongest of these satisfy all seven of Bovens’ requirements for relationships manifesting narrow accountability. For Bovens, a relationship qualifies as affording narrow accountability when: There is a relationship between an actor and a forum (1), in which the actor is obliged (2) to explain and justify (3) defined aspects of their conduct (4), where the forum can pose questions (5) and pass judgement (6), and where the actor may face consequences, whether formal or informal (7) [17]. All seven requirements enable the stakeholder to attempt to exert some leverage over the actor’s behaviour. Example mechanisms are listed in Figure 1, e.g. internal audit.%, with a longer working list and references to further reading available by contacting the authors.

A key point of the ecosystem perspective is materiality and the emergence of an appropriate level of accountability. Different aspects of AI have very different levels of associated risk in terms of negative outcomes, and hence different levels of materiality. Lower risk areas need fewer accountability mechanisms. Nonetheless, access to a broad ecosystem of mechanisms means that areas that suddenly increase in (perceived) risk have a number of options that can be scaled up or down until an appropriate level of accountability is obtained. A second key point is the overlap and mutual strengthening that is possible when mechanisms between layers work together. For example, corporate financial fraud is a highly material risk, motivating the introduction of a continually-evolving, broad-based ecosystem of accountability mechanisms so that the economy can continue to function [67]. Example areas of mutual reinforcement among mechanisms include public scrutiny, via whistleblowing and investigative journalism, providing a route to surface fraud when internal mechanisms are insufficient or have become captive to fraud, which drives better mechanisms and greater trust at all layers including the corporate layer. At the same time, standardised reporting in public accounts and transparency initiatives support public scrutiny and civil society to analyse trends and issues. External audits and credit rating agencies provide reassurance to market participants with access to internal audit providers drawing on the same expertise  as the external auditors. In the case of AI systems, when it comes to transparency, much of the recent emphasis has been placed on the creation of explainability tools. In an accountability ecosystem, however, it should be recognised that different explainability tools may be needed to provide transparency at the right level of abstraction for the different stakeholders, from data scientists to senior management and the general public.  

The ecosystem lens is attractive as it does not specify a single mechanism that has to be present. Ecosystems flex and evolve over time to develop new mechanisms and adjust to what is available and where materiality is highest. Nonetheless, the lens allows us to identify layers with weak mechanisms, consider the possible consequences of this and the likely behavioural responses across the stakeholder layers, and suggest areas where introducing or strengthening mechanisms might help bring an ecosystem into a stable state at which there is reduced risk and therefore reduced cost. Such a risk-based approach needs to recognise the priority areas for accountability in AI, as discussed next.

\subsection{Priority areas to develop AI accountability}
The foregoing discussion suggests that the majority of AI accountability activity has been taking place to date at the civil society stakeholder layer, noting that activity at other layers is likely larger than it appears, being less public by design. Government and regulatory mechanisms are also under active consideration [77]. The gulf between published principles and provable practice has likely exacerbated concerns. Fledgling attempts in the other layers, such as the ethical boards and recruited experts of Google or the Civil Rights Audit of Facebook, have failed to generate positive traction. An ecosystem perspective provides one possible contributing explanation for the difficulties they faced: these attempts were formally initiated in the corporate layer, but were not seeking to support the board or managers. Instead, they were directed at mollifying an external audience. 

These external facing exercises, especially Facebook’s audit, took place prematurely, opening processes to external scrutiny before they had been subject to internal scrutiny to identify and begin acting on areas of known weakness. The exercises were broad in their scope, addressing large swathes of operations, rather than focusing on specific areas of the business. %From another perspective, they were also too limited in their scope, in that they held individual companies accountable to an external, platonic ideal. 
A more constructive route is benchmarking: comparing companies against each other (in this case, big tech companies), identifying which ones are more robust in which areas, promoting good practice and incremental improvement, recognising that no one current actor is demonstrably ethically flawless but would suffer unduly if disclosing details unilaterally. 

Our suggestion is that stronger mechanisms in the corporate layer are needed before boards and senior managers can engage constructively with external challenge. For instance, rather than an external auditor delivering on a large scope for an external audience, the principles of internal audit suggest appointing an external expert to individually narrow scopes across a broad, multi-year programme of work, with the purpose of reporting to internal management and improving processes. Just as the value of external financial audits increases with robust, prior internal audits, the same can be true of accountability. This serves to connect the different subsets of AI accountability and operationalisation of AI principles to the risk management processes that are commonplace in large firms. It has the advantage of placing AI-related risks in the context of other risks, so that they can be prioritised and mitigated accordingly, using a range of pre-existing strategies.

The external firms commissioned to provide internal audit services, whether expert or general, seek to operate across a range of different companies. Some audit and technology firms have begun developing services for AI [68, 69] but further uptake, process development and maturation is needed before they will be able to meet the ambition of standardised process audits in AI. Such an approach, when mature, can provide a route into good practice dissemination and benchmarking. Crucially, by building a base of industry-level benchmarking, entire sectors can follow the external accreditation model, which has been proven effective in other domains with limited regulation (e.g. the Investors in People Award in the UK for ethical Human Resources and the Fairtrade labels for ethical supply chain management). This in turn begins to flesh out accountability mechanisms available to market participants and strengthen the bridge between the corporate stakeholder layer and civil society. 

This perspective concurs with other researchers calling for more operationalisation of principles, risk management approaches and internal audit, e.g. [5, 12]. It also supports the nascent technologies, standards, and service offers emerging in this space. For instance, [59] reports that IEEE launched an Ethics Certification Program for Autonomous and intelligent Systems (ECPAIS) in 2018, aiming to create specifications for certification and processes for advancing transparency and accountability, although emphasising that this remains work in progress and not fully ready for industrial applications [70].  

If successful, this approach allows a diversity of tailored mechanisms (such as explainability tools, process checklists and standardised bias reporting) to be explored in a guided, voluntary manner, out of which those proving to be the most effective might be targeted in future regulation to be rolled out as mandatory. However, to be effective in time to respond to civil society concerns and regulatory pressure, these activities need to accelerate considerably from the current level of engagement. In what follows, we illustrate one particular subset of the ecosystem approach with respect to AI in the gambling industry. 

\section{AI Mechanisms in the Gambling Industry}

The authors of this paper have been supporting Playtech plc advance its AI programme, including the use of supervised ML models to identify online gamblers at risk of harm [71]. The ML models seek to predict when changes in a player's interactions with the gambling platform, such as the intensity, frequency or volatility of betting behaviour and financial deposit and withdrawal transactions, might indicate a propensity for a player to be experiencing harm, as self-declared by the player. Since 2019, a key strand of Playtech's R\&D has been supporting accountability and ethical application of ML systems estimating player-level risk. In this section, we describe an approach to operationalising ethical principles and to support explainable AI at different stakeholder levels. We provide a worked example for future initiatives to build upon, which illustrates industry-led mechanisms playing an important role within a broad accountability ecosystem. 

We first explain how general ethical principles and long lists of potential issues need to be localised and prioritised for specific industry use cases in order to specify guidelines, processes and key performance indicators (KPIs). Second, we report specific technical investigations that define, identify and intervene in the ML system in order to address an example aspect of the ethics of AI, namely, avoiding unfair bias, and a separate strand of R\&D focused on explainable AI. Finally, we discuss how these activities might be developed in the future and interrelated with other mechanisms in a broad and effective accountability ecosystem.

\subsection{Localisation of general ethical principles:}
For a new organisation considering principles to adopt, there is no shortage of pre-existing language and frameworks to draw on. Different researchers synthesising sets of published principles have distilled 11 principles [44], five principles supported by 22 system requirements (c.f. [11]), and 10 themes with 50 detailed principles [12]. In order to operationalise these principles for a particular organisation and make progress with tangible reforms, it is important to localise general statements onto the relevant sector and prioritise the list of technologies in current use or under active development in that sector.

Working with the five high-level principles (beneficence, non-maleficence, autonomy, justice and explicability) and corresponding system requirements from [11], AI tools to support the online gambling industry have greater emphasis on “justification” and “stakeholder participation” than being e.g. “environmentally friendly” within the beneficence principle, since the training data sets and ML models in current practice are not large by comparison e.g. with language models such as GPT-3 or image recognition systems. Similarly, system requirements like “fallback plan and general safety” have been written more with robotics, safety critical systems and autonomous physical entities in mind (e.g. Tesla’s crash risk liability system and other irreversible outcomes), rather than the narrow input-output ML applications that support the gambling sector today. 

Following a review by gambling sector experts within the Playtech group, a draft set of nine industry-specific guidelines were developed, with a mapping to each of the five high-level principles (Table 1). Whilst some of the guidelines remain generic and relevant across wider industries, localisation of principles have a particular emphasis on reducing gambling related harm, the largest negative externality associated with the gambling industry.

%Include Table here:
\begin{table}[ht]
\centering
\resizebox{\textwidth}{!}{
\begin{tabular}{ |c|c| } 
 \hline
 Guidelines & High-level Principles  \\
 \hline
  1 Invest in AI for responsible gambling to protect the vulnerable & Beneficence  \\ 
 2 Embrace explainability in sensitive applications of AI	 & Explicability \\
 3 Build ‘human-in-the-loop’ into AI systems where appropriate	 & Autonomy  \\ 
 4 Leverage AI to deliver entertainment, however, change products where evidence points towards harm	 & Beneficence \\
 5 Avoid creating or re-enforcing unfair biases	 & Justice  \\ 
 6 Be open about AI blind spots and failures	 & Justice \\
 7 Be scientifically robust and continually evaluate	 & Non-Maleficence  \\ 
 8 Incorporate security, privacy, and diversity by design	 & Non-Maleficence \\
 9 Empower all stakeholders, including customers, staff and Boards, in the possibilities and risks of AI	 & Beneficence  \\ 
 \hline
\end{tabular}}
\caption{Proposed gambling industry-specific ethical guidelines mapped to the five high-level ethical principles from [11]: beneficence, non-maleficence, autonomy, justice and explicability.}
\end{table}

In guideline 1, consumer protection is a requirement of all regulated gambling markets and AI can help not only meet existing requirements but drive innovations that protect consumers from gambling-related harm. This includes using AI to develop systems that can classify or predict whether a person may be at risk of harm. Guideline 2 takes a generic principle, explainability, and applies it to the most sensitive applications of AI in the gambling industry, such as risk prediction systems. Furthermore, we would expect explainability to be measurable, with global model explainability enabling third-party evaluation of how such systems work (e.g. by a regulator or scientific experts), and with local explainability enabling specific explanations that can be used to inform customers on how they can change their gambling behaviours to decrease their risk of harm.

In guideline 3, the relevance of ‘human in the loop’ is associated to those processes and scenarios where automated decision making can have a material impact on the customer, e.g. account closure. In such circumstances, a machine may not have the full context to make the right decision. Guideline 4 highlights that the capabilities of AI are often utilized to not only help build entertaining gambling products, but also in marketing products and content that help improve the consumer experience, e.g. recommending services that a consumer may enjoy. However, if there is evidence to suggest that a product or service could be contributing towards harm, such as a product design or marketing technique encouraging excessive consumption, such a product or service should be changed or removed as a precautionary measure until otherwise proven not to be harmful. To enable gambling operators and suppliers to commit to this principle, they must invest in developing processes that include supporting post-release product evaluation, as well as considering industry specific product guardrails such as automatically limiting an AI system’s ability to market products to consumers who show certain signs of risk.

In what follows, we shall focus on processes that have been developed and applied to the gambling industry to assess wider industry concerns: algorithmic bias and potential AI blind spots (guidelines 5 and 6) and the requirement for explainability (guideline 2). The remaining guidelines (guidelines 7 to 9) refer to best practices that can be applied generally across all industries with an emphasis on continued training in the face of developments in AI, safety and the recognition of the value of diversity across the AI programme. For instance, diverse groups, including individuals with lived experience of gambling and gambling harms, need to be drawn upon for technology development, trust-enhancing mechanisms and oversight to function well. Data set design and data gathering protocols can also encode social biases to the detriment of technology performance. Considerations of social diversity and diverse teams need to be explicitly and sufficiently engaged [76]. Similarly, the unfair outcomes referred to in guideline 5 might be unfair across certain socio-demographic categories, as noted in [11] and discussed further below. 

\subsection{Avoiding unfair bias}
Once industry-specific guidelines have been drafted, specific processes and measures can be developed. This subsection provides a case study on the guideline “avoid creating or re-enforcing unfair bias”. This guideline was converted into an audit using a structured six step process: (1) prioritise scope, (2) prioritise bias categories, (3) define bias metrics, (4) analyse bias presence, (5) form and implement a plan given overall estimated costs/benefits, and (6) monitor and reflect. Each step is explained below, describing its application in our case study of the online gambling industry, with further details available in [19].

\textbf{Step 1: Prioritise scope}. Identify and list the complex algorithms within the organisation, specifying and justifying which are in scope as having the potential for bias to lead to material harm and which are excluded for now. For instance, in this case study, we prioritised a supervised machine learning model classifying the risk of problem gambling for individual players, which relates directly to individual user-level harm and increasing regulatory requirements in a number of jurisdictions.

\textbf{Step 2: Prioritise bias categories}. For each algorithm in scope: develop, maintain, and periodically review a list of categories within which there may be a concern about bias, such as players’ self-reported gender categories, different levels of customer status (beginner vs experienced players), protected characteristics and so on. Relevant categories should be listed even if not currently available for analysis (e.g. in our use case, data are not collected about player ethnicity). By being listed, it supports future conversations about whether ad-hoc data collection might be warranted or issues monitored in other ways (e.g. via customer service interactions). We first prioritised gender given concerns in the research literature that problem gambling was increasing among women, that gambling-related data sets were more likely to identify patterns relevant to male players, and given that GDPR-related caution by the industry have caused the gender variable to have been removed from deployed ML models.

\textbf{Step 3: Define bias metric(s).} Having identified the category within which we wanted to examine for bias, we decided where the bias might be observed and what statistical tolerance threshold should be applied for determining bias requiring action.\footnote{There are limitations in the risk outcome labelling, which relate more directly to other ethical principles (e.g. data integrity as part of non-maleficence) and are subject to a separate R\&D strand.} We looked at predictive accuracy by gender, reflecting a concern that a gender category might have disproportionately many false alerts or missed positives. Since chance variations mean that achieving exactly equal performance across categories cannot be expected, in this exploratory study, we have set a threshold of +/- 2\% of True Positive Rate (TPR) and True Negative Rate (TNR), reflecting the variation in performance obtained from a ten-fold cross-validation on the original ML model.

\textbf{Step 4: Analyse bias presence.} The bias metrics should now be evaluated quantitatively against each category of interest. A variety of tools and techniques are available depending on the type of bias metric chosen, ranging from descriptive data on group proportions, chi-squared test to compare against a benchmark community, analysis of model performance changes with variable inclusion/exclusion, and use of explainable AI techniques when the variable in question is included in the ML model. 

In this case study, gender data were voluntarily provided by a subset of players, available on two data sets for a Bingo-focused brand and a Casino-focused brand, shown in Table 2 as self-reported male (M), self-reported female (F), and unspecified (U). We first checked whether gender might be influencing model outcomes through its correlation with other included variables, using model-matched indirect identification, as explained in [19]. The study concluded that there is little potential for indirect identification of gender in the model (the ML model, a Random Forest (RF), is unable to predict gender as an outcome variable). The primary bias metric, as defined in step 3, is shown in Table 2. We see that differences by gender are within the specified tolerance threshold for Operator 1 but are exceeded for Operator 2, where the model is performing better for women than for men. This evaluation prompted a value-alignment question on the value of positive action for further discussion.

\begin{table}[ht]
\centering
\resizebox{\textwidth}{!}{
\begin{tabular}{ |c|c|c| } 
 \hline
Possible metric by: (F)emale / (M)ale / (U)nspecified &
	Operator 1 
(slots-focused, n=4,340) &	Operator 2 
(Bingo-focused, n=18,275) \\
\hline
Gender balance in training data &	F: 20.6\% M: 32.6\% U: 46.8\%	& F: 36.5\% M: 10.4\% U: 53.1\% \\
Self-exclusion outcome & F: 20.4\% M: 24.4\% U: 16.8\% & F: 17.1\% M: 18.7\% U: 22.3\% \\
Baseline RF model TPR & F: 67.0\% M: 65.3\% U: 66.5\% & F: 53.7\% M: 46.5\% U: 52.9\% \\
Baseline RF model TNR & F: 94.4\% M: 95.1\% U: 95.0\% &	F: 96.7\% M: 98.1\% U: 97.2\% \\
Baseline RF model accuracy & F: 88.8\% M: 87.9\% U: 90.2\% & F: 89.3\% M: 88.5\% U: 87.4\% \\
\hline
\end{tabular}}
\caption{Gender metrics from two gambling operators: The machine learning model is a Random Forest (RF); n denotes the total number of players evaluated per operator. The table shows the percentages of predicted self-exclusion as a proxy for harm, true positive rates (TPR), true negative rates (TNR), and model accuracy per gender for each operator.}
\end{table}

\textbf{Step 5: Form and implement a plan.} Results from step 4 can be evaluated against thresholds set in step 3 to decide on the urgency for action, the costs and trade-offs that might be involved in changing the model. Various methods exist for adjusting for bias in a fast-evolving field which precludes the recommendation of a single ML approach (see [73] for a taxonomy of nine possible methods, based on making adjustments either at the input level, the model level or the output level). 

In this case study, we sought to reduce the gender disparity in Operator 2 models without requiring an intrusive obligatory collection of gender data from players. Two methods were tested: (1) re-instating gender as an input variable in the ML model; (2) an ensemble blind-separate model approach, in which three separate models are trained from male-only, female-only and unspecified-only data sets, test players are run through each model (thus being blind to the test player’s gender) and the highest risk classification is adopted, treating TPR as the priority metric as part of a harm reduction initiative. The first method did not succeed, reducing gender disparity only by worsening TPR for women. However, the second method reduced the TPR gender disparity from 7.2\% to 4.0\%, a reduction of 44\%, with a small loss in overall accuracy. %At this stage, this improvement on one data set is not sufficient to implement operational changes given the other costs and consequences involved, but the topic remains under review with further sector discussions planned for 2021.

\textbf{Step 6: Monitor and reflect.} Any process to address algorithmic bias is likely to have limitations in its scope and approach. These limitations should be documented and monitored over time with further gathering of data on system performance. %In this case study, this step remains live in 2021, with industry discussions currently being scheduled.
From a methodological perspective, we are conscious of limitations around partial, voluntarily-provided gender data, the self-reported outcomes used as the labels for training the models, the complex and changing notion of fairness, the difficulty in identifying a benchmark for the right balance of outcomes in our use case, and the need for a broader sector engagement to arrive at judgements that can be shown to have been deemed acceptable. 

The importance of following a structured process lies in the ability to recognise the importance of all decision steps in the process, from setting the scope and metric through to reflection. Teams can then consider who to engage to support different decisions. For instance, the prioritising of bias categories and setting of tolerance thresholds might be best developed as part of cross-sector consultation, ensuring adequate representation of diverse groups and lived experience. By documenting the process throughout, it becomes transparent and auditable. Unless zero bias in all areas could be achieved, being able to explain the process and its decision points is key to enabling effective internal and external scrutiny. In this way, it is possible to build trust that the ethical principle is being duly upheld, or to produce constructive suggestions for how it might be improved.

\subsection{Addressing AI black-box explainability:}
A high-level process for implementing guidelines around explainability can follow a similar logic as the six-step process for algorithmic bias seen earlier: (1) prioritise scope for explainability – what does the relevant audience want to understand about the model and for what purpose? (2) select an explainability technique that supports the scope; (3) conduct analysis; (4) discuss results with domain experts and audience representatives as applicable; (5) form and implement a plan given overall estimate of cost/benefit; (6) monitor and reflect. As with algorithmic bias, the benefit of a process is that it can be documented, reviewed, challenged and improved upon, both to build trust in the system and to identify areas to do more.

The many aspects of explainable AI (XAI) come alive as part of the proposed accountability ecosystem in the case of the Playtech case study. Playtech develops ML models typically for third-party operators (licensees) to implement as embedded software solutions. In addition to the data scientists who are part of the development team and may wish to learn from XAI to improve system performance, and the different operators who may be interested in understanding the added value of a responsible gambling platform, XAI is relevant to the regulator (e.g. the Gambling Commission in the UK) who may be interested in obtaining evidence of how the system is used to protect the vulnerable, and ultimately to the player (the end customer) who has a "right to explanation" in the context of GDPR, and who may be keen to know why the system might have selected them as being potentially at risk or why the system recommended that they should take a break for a certain period of time. Such a rich ecosystem highlights the need for different explainability tools for different stakeholders and their specific purposes. Further stakeholders include internal management at the company, customer service and civil society actors (such as the UK GamCare, an independent charity that provides support for anyone affected by problem gambling). Finally, XAI tools may be needed to support the operationalisation of principles other than those associated with transparency alone, inasmuch as \emph{opening the black box} may be needed for the implementation of those principles, such as fairness. Although it has been argued that ML experts have chosen to focus on explainability because it is "the easy choice" compared to the difficult choices of ethics in AI and stakeholder engagement [44], XAI with all its facets is likely to become a precondition for much that is needed to follow, and as such it is of fundamental importance. 

\begin{figure*}
  \centering \includegraphics[width=0.8\textwidth]{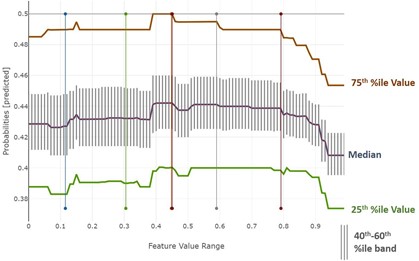}
  \caption{Feature risk curve for night-play versus problem gambling: The curve describes how a supervised ML model, in this case a Random Forest, interprets the change in percentage of a set of players' gambling during night-time hours (“night-play”) from 0\% to 100\% of their total play (x-axis) w.r.t. the probability that a player should be predicted by the model as being at risk of problem gambling according to a balanced data set (y-axis) with all other model features being equal.}
\end{figure*}

We now exemplify how risk curves are used by Playtech as part of its XAI toolkit (c.f. Figure 2). As mentioned, different tools and data visualization techniques are needed by different stakeholders, but every XAI approach needs to be measurable so as to provide an adequate level of confidence in the soundness and relevance of an obtained explanation [80]. For this reason, many off-the-shelf XAI packages have been proved inadequate for Playtech, with XAI requiring in-house development, continual learning and a bespoke evaluation and validation of systems by data scientists working alongside domain experts. Being able to describe how an ML model interprets the risk levels corresponding to a given input variable helps licensees understand and improve trust in the system, as well as identify possible blind spots for future model improvement. It is also important for customer service staff training and interaction with players about their risk scores. As well as explaining the model’s average risk scores corresponding to different values of individual input features (Figure 2), it is important to be able to describe the uncertainty around such values - XAI that \emph{knows when it does not know} - not least to try and account for the interactions that may exist between the many input features which is very difficult to describe in full. 

Feature risk curves [18] can be seen an a feature-specific global explainability method with high fidelity by design. Similar to the Explainable Boosting Machine (EBM) [72] derived from Generalized Additive Models (GAMs), feature risk curves treat any supervised ML model as an oracle, but apply to classifiers rather than regression models. Given a set of examples (i.e. the experience of multiple players), the curves describe the range of classification probabilities obtained for those players if the value of a feature of interest is set sequentially to each of its 100 percentile values. For this reason, the method is global (it does not seek to explain the experience of a single player) but feature-specific (it discounts feature interactions for the sake of simplicity, although the addition of interaction terms is possible and part of work in progress). The method is said to have high-fidelity by design because the risk curves are obtained by querying the model directly, thus making the curves accurate w.r.t. the underlying ML model.   

%The resulting curves show, with an indicative confidence interval, how the model interprets changes in the feature of interest, holding all other variables constant. The graphic display is similar to EBM-derived variable-response curves [72], with the benefit of being applicable to any ML classifier not just generalised additive models, but the disadvantages of representing a simplification of the feature’s role in the model (i.e. input variable interaction effects translate into the confidence interval size but are not explicitly visible) and not allowing a direct in-model intervention for correcting any identified bias.

We have applied feature risk curves to the Random Forest models studied in [71]. The graph in Figure 2 describes only one such curve as part of the analysis of the influence of night-play on the model's prediction of harm. The curves were shown useful to stimulate discussion with sector experts. In this particular example, sector experts have challenged the reduction in risk shown at the high end of the curve. The model indicates less of a concern about the players who gamble almost exclusively at night. But what if the labels used by the system for model training - player self-exclusion - were insufficient for the most serious and intense players? This exemplifies a great benefit of EBM and feature risk curves: the ease with which AI non-experts can interact with and intervene in the curves in order to change an undesirable trend of the system. At present, the intervention approach for this blind spot is post-hoc, with gambling operators advised to monitor the few players with the most intense gambling activity, even if the model does not identify them as high risk. A possible future intervention might be to seek to adjust the model and re-evaluate it. This illustrates the six-step process in action, and how it may produce simplified models, improved models and promote continual learning.

\section{Discussion}

Although the research programme described above supports a small part of the accountability ecosystem, it can play an important role in the ecosystem towards process standards and transparency. When a structured, documented approach to algorithmic bias is supported by KPIs, it can be applied to a broad set of bias categories beyond just gender, and to a broad set of priority algorithms in the gambling sector beyond problem gambling. It offers an internal accountability mechanism within the corporate layer, enabling oversight, mitigation and an understanding of the remaining risks. Such an understanding should trigger active efforts to mitigate risk and improve practice. It should then invite external experts to challenge the work, surface further areas for improvement, and share the results with the wider sector, whether as named case studies or as anonymised benchmarks to be drawn on by industry mediators such as external accreditation bodies or audit providers. This process can result in published reports, both at the individual actor level and industry-level. This extra information can then be reviewed, analysed and challenged by civil society and government. Where a particular stakeholder considers the approach to be insufficient, they now have a process within which they can specify what improvements to ask for, such as in the gender bias example, a stricter tolerance for bias, a different metric for measuring bias, etc. Debate and challenge can proceed more constructively, even if just as passionate and urgent as before. 

Similar processes and KPIs can be identified for the other industry guidelines suggested for the gambling sector. Debate can also focus on whether these are the necessary and sufficient guidelines for the sector, how values might be specified that underpin certain guidelines, and how trade-offs between values might be managed in practice. Some specific processes or KPIs might later be mandated through regulation, to ensure wider adoption and more robust scrutiny as a result of lessons learned.

From where we are today, a greater level of formalisation and consistency around transparency, standards, expectations and liability will help AI become more accountable and help practitioners defend themselves as having acted appropriately. However, we should be on guard against introducing a "checkbox mentality" as a result of over-complicated procedures. If not balanced out by other mechanisms or a culture of openness, such a mentality can reduce the free thinking, creativity and challenge around new ideas.

\section{Conclusion and Future Work} 
This paper has set out the high tempo of concern among diverse civil society members around the ethical behaviour of AI teams, with evidence from news flow, monitoring units and public polling, and described the resulting rapid progression towards regulatory intervention. We have presented an accountability ecosystem perspective in which civil society pressure and regulatory intervention work best in conjunction with other accountability mechanisms. These include standardised processes, KPIs, senior management visibility, and internal audits alongside market participant mechanisms like external audits and eventually accreditation. 

At this level of operationalisation, industry or organisation-specific approaches become necessary, and we shared the work-in-progress of R\&D on accountability within the online gambling industry, including step-by-step processes for addressing algorithmic bias and model explainability. Far more is needed to be built on this to create the systems and processes to drive a robust accountability ecosystem in AI. In fact, procedures and mechanisms can only go so far in building trust in AI which also requires a broad-based collective understanding, including education, a culture of transparency, and improved technologies for embedding ethical decision-making into AI systems [75]. 

The aspirations of regulators are high, perhaps even out-of-sight given current technology and practices. The Information Commissioner's Office (ICO) in the UK expects that: \emph{there should be no loss of accountability when a decision is made with the help of, or by, an AI system, rather than solely by a human. Where an individual would expect an explanation from a human, they should instead expect an explanation from those accountable for an AI system} [77]. However, if we collectively fail to make progress towards these goals, towards measurable XAI and implementing these accountability mechanisms, we can look forward to immature regulation operating in an unbalanced ecosystem, being either ineffective or overwhelming, and continued frustration from a civil society that lacks constructive channels through which to direct their anger and to help shape AI practice.

% \begin{figure}[!htbp]
%   \centering
%   \includegraphics[width=1\hsize]{filename.eps}
%   \caption{caption} \label{fig:label}
% \end{figure}

\section*{Acknowledgements}
Thanks to Playtech Plc for its R\&D programme to improve responsible gambling and AI accountability, and for making data, domain expertise and software implementations available to support this work.

\section*{References} 

{\small
\noindent [1] Asimov, I. (1950). I Robot. Gnome Books.

\noindent [2] Automating Inequality: How High-tech Tools Profile, Police and Punish the Poor, Virginia Eubanks, St. Martin's Press, New York, NY, 2018.

\noindent [3] He got Facebook hooked on AI. Now he can't fix its misinformation addiction, \url{https://www.technologyreview.com/2021/03/11/1020600/facebook-responsible-ai-misinformation}, MIT Technology Review. Accessed: 2021-03-28.

\noindent [4] European Commission. (2020). White paper on Artificial Intelligence: a European approach to excellence and trust. 

\noindent [5] Canca, C. (2020). Computing Ethics: Operationalizing AI Ethics Principles. Commun. ACM 63, 12 (December 2020), 18–21. DOI:\url{https://doi.org/10.1145/3430368}.

\noindent [6] Daniel Zhang, Saurabh Mishra, Erik Brynjolfsson, John Etchemendy, Deep Ganguli, Barbara Grosz, Terah Lyons, James Manyika, Juan Carlos Niebles, Michael Sellitto, Yoav Shoham, Jack Clark, and Raymond Perrault, The AI Index 2021 Annual Report, AI Index Steering Committee, Human-Centered AI Institute, Stanford University, Stanford, CA, March 2021.

\noindent [7] CDEI. (2020). AI Barometer Report: June 2020. London: Centre for Data Ethics and Innovation, UK.

\noindent [8] Fernandes, Pedro M., Santos, Francisco C., and Lopes, Manuel (2020). Norms for Beneficial A.I.: A Computational Analysis of the Societal Value Alignment Problem. AI Communications, vol. 33, no. 3-6, pp. 155-171, 2020.

\noindent [9] \url{https://digital-strategy.ec.europa.eu/en/library/proposal-regulation-\\laying-down-harmonised-rules-artificial-intelligence-artificial-intelligence?s=09}

\noindent [10] National Audit Office (2014). Using alternatives to regulation to achieve policy objectives. London: NAO.

\noindent [11] Morley, J., Floridi, L., Kinsey, L.,  Elhalal, A. (2020). From What to How: An Initial Review of Publicly Available AI Ethics Tools, Methods and Research to Translate Principles into Practices. Science and Engineering Ethics (2020) 26:2141–2168, \url{https://doi.org/10.1007/s11948-019-00165-5}.

\noindent [12] Clarke, R. (2019). Principles and business processes for responsible AI. Computer Law \& Security Review, Volume 35, Issue 4. 2019, Pages 410-422. \url{https://doi.org/10.1016/j.clsr.2019.04.007}.

\noindent [13] Smit, Koen; Zoet, Martijn; and van Meerten, John. (2000). A Review of AI Principles in Practice. PACIS 2020 Proceedings, 198. \url{https://aisel.aisnet.org/pacis2020/198}.

\noindent [14] Whittlestone, J. et al (2019). The Role and Limits of Principles in AI Ethics: Towards a Focus on Tensions. AIES’19, January 27–28, 2019, Honolulu, HI, USA.

\noindent [15] Goodman, B. (2016). A Step Towards Accountable Algorithms? Algorithmic Discrimination and the European Union General Data Protection. 29th Conference on Neural Information Processing Systems Information Processing Systems (NIPS 2016), Barcelona, Spain, December 2016.

\noindent [16] Nathalie A. Smuha (2021): From a race to AI to a race to AI regulation: regulatory competition for artificial intelligence, Law, Innovation and Technology. DOI:10.1080/ 17579961.2021.1898300

\noindent [17] Bovens, Mark. (2006). Analysing and Assessing Public Accountability. A Conceptual Framework. European Governance Papers (EUROGOV) No. C-06-01, \url{http://www.connex-network.org/eurogov/pdf/egp-connex-C-06-01.pdf}.

\noindent [18] S. Dragicevic, A. d’Avila Garcez, C. Percy, and S. Sarkar.  Understanding the risk profile of gambling behaviour through machine learning predictive modelling and explanation. In NeurIPS Workshop Knowledge Representation meets Machine Learning, KR2ML’2019, Vancouver, Canada, December, 2019.

\noindent [19]  C.  Percy,  A.  d’Avila  Garcez,  S.  Dragicevic,  and  S.  Sarkar. Lessons learned from problem gambling classification: Indirect discrimination and algorithmic fairness. In Proc. AAAI Fall Symposium, AI for Social Good, Washington DC, USA, November, 2020.

\noindent [20] Ledford, Heidi (2019). Millions of black people affected by racial bias in health-care algorithms. Nature. 574 (7780): 608–609. doi:10.1038/d41586-019-03228-6.

\noindent [21] Kleinberg, J., Ludwig, J., Mullainathan, S., Rambachan, A. (2018). Advances in big data research in economics: Algorithmic fairness. AEA Papers and Proceedings 2018, 108: 22–27 \url{https://doi.org/10.1257/pandp.20181018, 2018}.

\noindent [22] Roach, J. (2018). Microsoft improves facial recognition technology to perform well across all skin tones, genders. (26 June 2018), \url{blogs.microsoft.com}. 

\noindent [23] AI is sending people to jail—and getting it wrong. MIT Technology Review. \url{https://www.technologyreview.com/ 2019/ 01/21/137783/ algorithms-criminal-justice-ai/}

\noindent [24] Haag, Rechtbank Den (6 March 2020). ECLI:NL: RBDHA:2020:1878, Rechtbank Den Haag, C-09-550982-HA ZA 18-388 (English). \url{uitspraken.rechtspraak.nl} (in Dutch).

\noindent [25]  Simonite, T. (2020). Skewed Grading Algorithms Fuel Backlash Beyond the Classroom. (19 Aug 2020). \url{Wired.Com}. 

\noindent [26] \url{https://igamingbusiness.com/veikkaus-\\sets-out-ethical-guidelines-for\\-artificial-intelligence/}. 

\noindent [27] Discussions at Digital Summit hosted by SBC in September 2020. 

\noindent [28] Crawford, Kate, Roel Dobbe, Theodora Dryer, Genevieve Fried, Ben Green, Elizabeth Kaziunas, Amba Kak, Varoon Mathur, Erin McElroy, Andrea Nill Sánchez, Deborah Raji, Joy Lisi Rankin, Rashida Richardson, Jason Schultz, Sarah Myers West, and Meredith Whittaker. AI Now 2019 Report. New York: AI Now Institute, 2019.

\noindent [29] \url{https://ainowinstitute.org/AI\_Now\_2018\_Report.html}. 

\noindent [30] Gibbs, Samuel (17 July 2017). "Elon Musk: regulate AI to combat 'existential threat' before it's too late". The Guardian. 

\noindent [31] Bostrom, Nick (2014). Superintelligence: Paths, Dangers, Strategies.

\noindent [32] \url{http://futureoflife.org/misc/open\_letter}. 

\noindent [33] Zhang, B. (2019). Public opinion lessons for AI regulation. The Brookings Institution. 
\noindent [34] Zhang, B., Dafoe, A. (2019). Artificial Intelligence: American Attitudes and Trends. (Jan 2019). Center for the Governance of AI, Future of Humanity Institute, University of Oxford. \url{https://governanceai.github.io/}.

\noindent [35] Committee on Standards in Public Life. (2020). Artificial Intelligence and Public Standards: public polling. (11 February 2020). \url{www.gov.uk}.  

\noindent [36] Boyon, N. (2019). Widespread concern about artificial intelligence (1 July 2019). Available from \url{ipsos.com}. 

\noindent [37] \url{https://www.hrw.org/topic/technology\\-and-rights}.

\noindent [38] \url{https://www.amnesty.org.uk/issues/technology-and-human-rights}.

\noindent [39] OECD. Guidelines on the Protection of Privacy and Transborder Flows of Personal Data. 

\noindent [40] \url{http://eur-lex.europa.eu/legal-content/EN/TXT/HTML/?uri=CELEX:31995L0046}.

\noindent [41] \url{https://eur-lex.europa.eu/eli/reg/2016/679/oj}.

\noindent [42] \url{https://siriuslegaladvocaten.be/en/gdpr-in-new-york/}. 

\noindent [43] \url{https://www.oecd.org/science/forty-two-countries-adopt-new-oecd-\\principles-on-artificial-intelligence.htm}. 

\noindent [44] Jobin, A., Ienca, M., Vayena, E. (2019). Artificial Intelligence: The global landscape of ethics guidelines. arXiv :1906.11668 [Cs].

\noindent [45] G20 Ministerial Statement on Trade and Digital Economy (2019). Tsukuba City, Japan: G20. 

\noindent [46] Babuta, Alexander; Oswald, Marion; Janjeva, Ardi (2020). Artificial Intelligence and UK National Security: Policy Considerations. London: Royal United Services Institute.

\noindent [47] State Council of the PRC's July 8. (2017). A Next Generation Artificial Intelligence Development Plan (State Council Document No. 35)

\noindent [48] Popova, Anna V., Gorokhova, Svetlana S., Abramova, Marianna G., Balashkina, Irina V. (2021), The System of Law and Artificial Intelligence in Modern Russia: Goals and Instruments of Digital Modernization, Studies in Systems, Decision and Control, Cham: Springer International Publishing, pp. 89–96.

\noindent [49] William D. Eggers, Mike Turley, and Pankaj Kamleshkumar Kishnani. (2018). The future of regulation: Principles for regulating emerging technologies. Deloitte. 

\noindent [50] Dean, T., Brown, R. (1995). Pollution Regulation as a Barrier to New Firm Entry: Initial Evidence and Implications for Future Research. Academy of Management Journal Vol. 38, No. 1  \url{https://doi.org/10.5465/256737}.

\noindent [51] Walker, J., Wellock, T. (2010). A Short History of Nuclear Regulation, 1946–2009 NUREG/BR-0175, Rev. 2. U.S. Nuclear Regulatory Commission. 

\noindent [52] Jorgensen, T. (2016). When energy drinks actually contained radioactive energy. The Conversation. 

\noindent [53] OECD. (2002). Reviews of Regulatory Reform: Regulatory Policies in OECD Countries From Interventionism to Regulatory Governance. 

\noindent [54] \url{https://www.vox.com/future-perfect/2019/4/4/18295933/google-cancels-ai-ethics-board}. 

\noindent [55] \url{https://www.bbc.co.uk/news/technology-55281862}. 

\noindent [56] \url{https://about.fb.com/wp-content/uploads/2020/07/Civil-Rights-Audit-Final-Report.pdf}. 

\noindent [57] \url{https://www.bbc.co.uk/news/technology-53333626}. 

\noindent [58] \url{https://about.fb.com/news/2021/01/roy-austin-facebook-vp-civil-rights/}. 

\noindent [59] J. Zhou, F. Chen, A. Berry, M. Reed, S. Zhang and S. Savage. (2020). A Survey on Ethical Principles of AI and Implementations, IEEE Symposium Series on Computational Intelligence (SSCI), Canberra, ACT, Australia, 2020, pp. 3010-3017.

\noindent [60] Celis, E., et al. (2019). A dashboard for controlling polarization in personalization. AI Communications, vol. 32, no. 1, pp. 77-89, 2019. DOI: 10.3233/AIC-180606.

\noindent [61] Van Belle S, Boydell V, George AS, Brinkerhoff DW, Khosla R (2018) Broadening understanding of accountability ecosystems in sexual and reproductive health and rights: A systematic review. PLoS ONE 13(5): e0196788. \url{https://doi.org/10.1371/journal.pone. 0196788}.

\noindent [62] Halloran, B. (2016) Accountability ecosystems: directions of accountability and points of engagement, Brighton: IDS Institute of Development Studies 2016.

\noindent [63] Chemonics. (2019). Accountability Ecosystems in Practice (Nov 2019).

\noindent [64] Halloran, B. (2015) Strengthening Accountability Ecosystems: A Discussion Paper.
Transparency and Accountability Initiative Think Piece, pp. 1-22.

\noindent [65] Maranke Wieringa. (2020). What to account for when accounting for algorithms: a systematic literature review on algorithmic accountability. In Proceedings of the 2020 Conference on Fairness, Accountability, and Transparency (FAT*20). Association for Computing Machinery, New York, NY, USA, 1–18. DOI:\url{https://doi.org/10.1145/3351095.3372833}. 

\noindent [66] \url{https://www.playtech.com/application/files/6816/1908/1881/Responsible\_Gambling\_Algorithms\_Roundtable\_1\_August\_2016\_FINAL.pdf}

\noindent [67] Wells, J. (2017). Corporate Fraud Handbook: Prevention and Detection. John Wiley \& Sons. 

\noindent [68] An Ethical Framework for Responsible AI and Robotics. \url{www.accenture.com/gb-en/companyresponsible-ai-robotics}.

\noindent [69] J. Samuel, KPMG launches framework to help businesses gain greater confidence in their AI technologies - KPMG Global, KPMG, Feb. 13, 2019.

\noindent [70] IEEE Launches Ethics Certification Program for Autonomous and Intelligent Systems, IEEE Standards Association, Oct. 02, 2018. \url{standards.ieee.org/news/2018/ieee-launches-ecpais.html}.

\noindent [71] C. Percy, M. França, S. Dragicevic, A. d’Avila Garcez. Predicting online gambling self-exclusion: an analysis of the performance of supervised ML models, International Gambling Studies, 16:2, 2016.

\noindent [72] R. Caruana, P. Koch, Y. Lou, M. Sturm, J. Gehrke, N. Elhadad. Intelligible Models for HealthCare: Predicting Pneumonia Risk and Hospital 30-day Readmission. KDD’15, Aug. 2015, Sydney, Australia.

\noindent [73] \url{https://tinyurl.com/4tw6w3pj}

\noindent [74] Claudio Corradini (Managing Director, Accenture) attending the Responsible Gambling Roundtable held at City, University of London in August 2016. Write-up available via \url{playtech.com}. 

\noindent [75] See for instance IBM's interviews on building trust in AI. \url{https://www.ibm.com/watson/advantage-reports/future-of-artificial-intelligence/building-trust-in-ai.html}.

\noindent [76] \url{https://aibusiness.com/document.asp?doc_id=760915}.

\noindent [77] \url{https://ico.org.uk/for-organisations/guide-to-data-protection/key-dp-themes/explaining-decisions-made-with-artificial-intelligence/part-1-the-basics-of-explaining-ai/definitions/}.

\noindent [78]  Cobbe, J., Lee, M., Singh, J. (2021). Reviewable Automated Decision-Making: A Framework for Accountable Algorithmic Systems. In Proceedings of the 2021 ACM Conference on Fairness, Accountability, and Transparency (FAccT'21). Association for Computing Machinery, New York, NY, USA, 598-609. 
\noindent [79] Kroll, J. (2021). Outlining Traceability: A Principle for Operationalizing Accountability in Computing Systems. In Proceedings of the 2021 ACM Conference on Fairness, Accountability, and Transparency (FAccT'21). Association for Computing Machinery, New York, NY, USA, 758-771.

\noindent [80] White, A., d'Avila Garcez, A. S. (2020). Measurable Counterfactual Local Explanations for Any Classifier. In Proceedings of the European Conference on Artificial Intelligence, ECAI 2020, Santiago de Compostela, Spain, August 2020. \url{https://ecai2020.eu/papers/514_paper.pdf}.

}

\end{document}